\documentclass[sigconf,natbib=true,screen=true,review=False,anonymous=False]{acmart}

\AtBeginDocument{%
  }

\makeatother

\setcopyright{acmcopyright}
\copyrightyear{2018}
\acmYear{2018}
\acmDOI{XXXXXXX.XXXXXXX}
\acmConference[Conference acronym 'XX]{Make sure to enter the correct
  conference title from your rights confirmation emai}{June 03--05,
  2018}{Woodstock, NY}
\acmPrice{15.00}
\acmISBN{978-1-4503-XXXX-X/18/06}

\usepackage{bm}
\usepackage{caption}
\usepackage{graphicx}
\usepackage{subfigure}
\usepackage{amsmath}
\usepackage{booktabs}
\usepackage{threeparttable}
\usepackage{multirow}
\usepackage{enumitem}
\usepackage{xcolor}
\usepackage{mathtools}
\usepackage{marvosym}
\usepackage[ruled]{algorithm2e}

\begin{document}

\title{CAME: Competitively Learning a Mixture-of-Experts Model for First-stage Retrieval}

\author{Yinqiong Cai}
\affiliation{
 \institution{CAS Key Lab of Network Data Science and Technology, ICT, CAS}
 \institution{University of Chinese Academy of Sciences}
 \city{Beijing}
 \country{China}
}
\email{caiyinqiong18s@ict.ac.cn}

\author{Yixing Fan}
\affiliation{
 \institution{CAS Key Lab of Network Data Science and Technology, ICT, CAS}
 \institution{University of Chinese Academy of Sciences}
 \city{Beijing}
 \country{China}
}
\email{fanyixing@ict.ac.cn}

\author{Keping Bi}
\affiliation{
 \institution{CAS Key Lab of Network Data Science and Technology, ICT, CAS}
 \institution{University of Chinese Academy of Sciences}
 \city{Beijing}
 \country{China}
}
\email{bikeping@ict.ac.cn}

\author{Jiafeng Guo}
\authornote{Jiafeng Guo is the corresponding author.}
\affiliation{
 \institution{CAS Key Lab of Network Data Science and Technology, ICT, CAS}
 \institution{University of Chinese Academy of Sciences}
 \city{Beijing}
 \country{China}
}
\email{guojiafeng@ict.ac.cn}

\author{Wei Chen}
\affiliation{
 \institution{CAS Key Lab of Network Data Science and Technology, ICT, CAS}
 \institution{University of Chinese Academy of Sciences}
 \city{Beijing}
 \country{China}
}
\email{chenwei2022@ict.ac.cn}


\author{Ruqing Zhang \\ Xueqi Cheng}
\affiliation{
 \institution{CAS Key Lab of Network Data Science and Technology, ICT, CAS}
 \institution{University of Chinese Academy of Sciences}
 \city{Beijing}
 \country{China}
}
\email{{zhangruqing, cxq}@ict.ac.cn}

\renewcommand{\shortauthors}{Yinqiong Cai et al.}
\newcommand{\modelname}[1]{CAME}

\begin{abstract}
The first-stage retrieval aims to retrieve a subset of candidate documents from a huge collection both effectively and efficiently. 
Since various matching patterns can exist between queries and relevant documents, previous work tries to combine multiple retrieval models to find as many relevant results as possible. 
The constructed ensembles, whether learned independently or jointly, do not care which component model is more suitable to an instance during training. 
Thus, they cannot fully exploit the capabilities of different types of retrieval models in identifying diverse relevance patterns.
Motivated by this observation, in this paper, we propose a Mixture-of-Experts (MoE) model consisting of representative matching experts and a novel competitive learning mechanism to let the experts develop and enhance their expertise during training. 
Specifically, our MoE model shares the bottom layers to learn common semantic representations and uses differently structured upper layers to represent various types of retrieval experts. 
Our competitive learning mechanism has two stages: (1) a standardized learning stage to train the experts equally to develop their capabilities to conduct relevance matching; (2) a specialized learning stage where the experts compete with each other on every training instance and get rewards and updates according to their performance to enhance their expertise on certain types of samples.
Experimental results on three retrieval benchmark datasets show that our method significantly outperforms the state-of-the-art baselines.
\end{abstract}



\maketitle

\section{Introduction}
\begin{figure}[!t]
\setlength{\abovecaptionskip}{-0cm}
\setlength{\belowcaptionskip}{-0cm}
\includegraphics[scale=0.44]{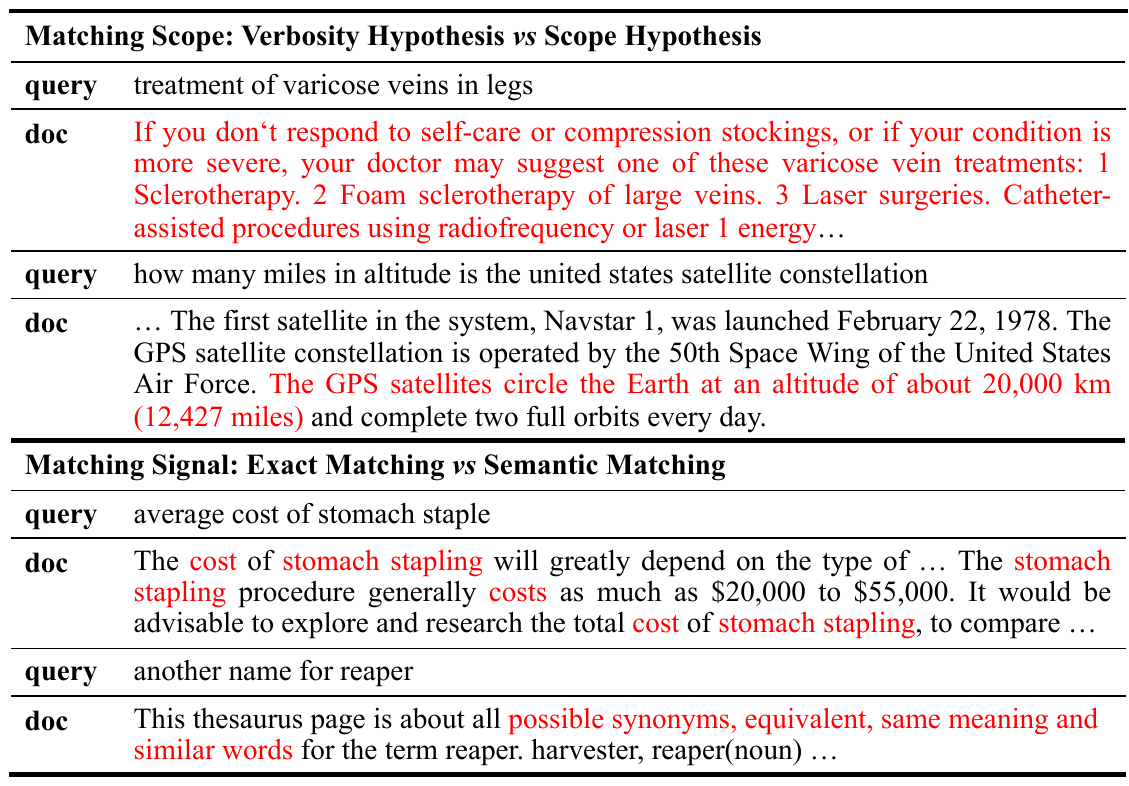}
\caption{Examples of relevant query-document pairs in MS MARCO~\cite{nguyen2016ms} to show diverse relevance patterns from different perspectives, e.g., matching scope and matching signal.}
\label{fig:patterns}
\end{figure}

The first-stage retrieval aims to retrieve potentially relevant documents from a large-scale collection efficiently, which is the foundation of various downstream tasks~\cite{guo2022semantic,lin2021pretrained}.
However, relevant documents have diversified matching patterns and identifying all of them is very challenging. For example, as shown in Figure~\ref{fig:patterns}, from the perspective of matching scope,  there are \textit{verbosity hypothesis} (i.e., the entire document is about the query topic, only with more words) and \textit{scope hypothesis} (i.e., a document consists of several parts concatenated together and only one of them is relevant)~\cite{jones2000probabilistic}.
From the perspective of matching signals, a document can be relevant to a query by either exact matching or semantic matching~\cite{guo2016deep}.

Aware of these different relevance patterns, various retrieval models have been proposed~\cite{karpukhin2020dense, formal2021splade, khattab2020colbert}.
In general, they can be categorized into three main paradigms:
(i) \textit{Lexical Retrievers}, e.g., DeepCT~\cite{dai2020context} and SPLADE~\cite{formal2021splade}, represent the document/query as a sparse vector of the vocabulary size, where the value of each dimension is the corresponding term weight. These approaches are inherently good at capturing exact matching.
(ii) \textit{Local Retrievers}, e.g., ColBERT~\cite{khattab2020colbert, santhanam-etal-2022-colbertv2} and COIL~\cite{gao2021coil}, conduct semantic matching between queries and documents at the term level, and measure the relevance according to the maximum similarity a query term can have to document terms. This type of method can identify the best local matching, which hereby is aligned with the scope hypothesis.
(iii) \textit{Global Retrievers}, e.g., DPR~\cite{karpukhin2020dense} and ANCE~\cite{xiong2020approximate}, represent the document and query as global-level low-dimension dense vectors and measure their relevance according to the similarity between the two vectors. They capture the global matching and fit in the verbosity hypothesis.
Since most retrievers have their focused relevance matching patterns, it is difficult for a single model to handle various matching needs and serve all queries well. This finding is also confirmed by the observations from earlier studies~\cite{gao2021complement, zhan2020repbert}: the retrieved results from different types of models have few overlaps, even when their overall quality is similar.

To address the issue that a single model is unable to fit all matching patterns, some work has proposed \textit{ensemble retrievers} to leverage the capabilities of multiple retrieval models~\cite{gao2021complement, kuzi2020leveraging, shen2022unifier}. 
A straightforward method is to train several retrievers independently with the entire training data and aggregate their scores to produce improved results~\cite{luan2021sparse, kuzi2020leveraging}.
However, the individual retrievers cannot interact sufficiently in such a way. 
A more advanced method is to learn component retrievers jointly with collaborative training mechanisms and aggregate their results.
For example, following the boosting technique, CLEAR~\cite{gao2021complement} and DrBoost~\cite{lewis2021boosted} train each component retriever with the residuals of the other retriever(s) of the different~\cite{gao2021complement} or same type~\cite{lewis2021boosted}; UnifieR~\cite{shen2022unifier} learns two retrievers and aligns one with the other based on their listwise agreements on the predictions.
Such ensemble retrievers have been shown to be very effective in improving retrieval quality and are widely employed in modern search engines~\cite{ling2017model}.
However, no matter learning the retriever with the samples that the other component retrievers perform well (i.e, UnifieR) or poorly (e.g., CLEAR and DrBoost) on, all of them could force the retriever to compromise to unsuitable samples (i.e., matching patterns) and result in sub-optimal overall performance.
Intuitively, it would be ideal to fully exploit the capabilities of different model architectures and make every component retriever especially focus on a certain type of sample.
Such a ``divide-and-conquer'' idea has been shown to be pivotal in the Mixture-of-Experts (MoE) framework~\cite{jacobs1991adaptive, shazeer2017outrageously}.

Motivated by the above, we build an MoE retrieval model consisting of multiple representative matching experts, and propose a novel mechanism to Competitively leArn the MoE model, named as \modelname{}.
Specifically, we include lexical, local, and global retrievers in a multi-path network architecture with shared bottom layers and top MoE layers. The shared bottom layers aim to extract common syntactic and semantic representations for all the experts. The MoE layers consist of three experts with different model architectures to capture pattern-specific features.
To guide the component experts in \modelname{} to specialize on certain types of samples, we competitively learn the MoE retriever in two phases. First, in the standardized learning stage, each expert is trained equally to develop the relevance matching capability and prepare for specialization. Second, in the specialized learning stage, the component experts compete with each other on every training instance, and they are trained proportionally to their relative performance among all the experts.
In this way, each sample is only used to update the experts that perform decently but has no or little impact on other experts.
During inference, each expert estimates the relevance score from its perspective to contribute to the final results.
Noted that, in contrast to the classical MoE models that pick experts spontaneously for each input sample, our method establishes more explicit connections between the experts and samples via their relative ranking performance.
From the above, we can readily urge each expert to fit the case it is skilled at and fully unleash the advantages of different model architectures.

We evaluate the effectiveness of the proposed model on three representative retrieval benchmark datasets, i.e., MS MARCO~\cite{nguyen2016ms}, TREC Deep Learning Track~\cite{craswell2020overview}, and Natural Questions~\cite{kwiatkowski2019natural}.
The empirical results show that \modelname{} can outperform all the baselines including various state-of-the-art single-model retrievers and ensemble retrievers significantly.
In addition, we conduct comprehensive analyses on how the model components and learning mechanism impact retrieval performance. It shows that employing multi-types of retrieval experts can capture diverse relevance patterns, and the competitive learning strategy is essential to facilitate the experts to learn their designated patterns, which together boost the retrieval performance. 
To sum up, our contributions include:
\begin{itemize}[leftmargin=*,topsep=1pt,parsep=1pt]
\item We propose a Mixture-of-Experts retrieval model that can orchestrate various types of models to capture diverse relevance patterns.
\item We propose a novel competitive learning mechanism for the MoE retriever that encourages the component experts to develop and enhance their expertise on dedicated relevance patterns.
\item We conduct extensive experiments on three representative retrieval benchmarks and show that \modelname{} can outperform various types of state-of-the-art baselines significantly.
\item We provide a comprehensive analysis on the model components, learning mechanism, and hyper-parameter sensitivity to better understand how they impact the retrieval performance.
\end{itemize}

\section{Related Work}
In this section, we review the most related topics to our work, including ensemble models for first-stage retrieval, and the Mixture-of-Experts framework.

\subsection{Ensemble Models for First-stage Retrieval}
The ensemble of several models has become a standard technique for improving the effectiveness of information retrieval (IR) tasks.
It has been studied extensively in the TREC evaluations~\cite{harman1995overview} and is the basis of Web search engines~\cite{ling2017model}.
Overall, there are two approaches to developing ensemble IR models~\cite{croft2002combining}, both with the motivation to improve retrieval performance by combining multiple evidences.
One approach is to create models that explicitly utilize and combine multiple sources of evidence from raw data~\cite{fisher1972general}. For example, existing studies have shown that combining information from multiple fields of the document can facilitate relevance estimation~\cite{fisher1972general}.
The other approach is to combine the outputs of several retrieval models by voting or weighted summation, where each prediction provides an evidence about relevance~\cite{fox1987architecture}.

Recently, with the development of pre-trained language models (PLMs), a variety of neural retrievers based on different relevance hypotheses have been proposed~\cite{guo2022semantic, zhao2022dense}.
Although these retrievers have achieved impressive performance on retrieval benchmarks~\cite{formal2021splade,hofstatter2021efficiently,zhou2022master,liu2022retromae}, their focuses (i.e., identifying relevance patterns) are distinct due to the differences in the model architecture. 
To combine the benefits from different retrieval models, the idea of building ensemble retrieval models has been explored by researchers~\cite{kuzi2020leveraging, gao2021complement, lewis2021boosted, shen2022unifier}.
A straightforward way is to train multiple retrievers independently and combine their predicted scores linearly to obtain the final retrieval results~\cite{luan2021sparse, kuzi2020leveraging, chen2021salient, lin2021densifying}.
Some work has been inspired by the boosting technique~\cite{gao2021complement, lewis2021boosted}. For example, \citet{gao2021complement} proposed CLEAR to learn a BERT-based retriever from the residual of BM25, and then combined them to build an ensemble retrieval model.
Also, the knowledge distillation technique is introduced into building ensemble retrievers, which tries to leverage the capability of companion retrievers to enhance the ensemble results~\cite{shen2022unifier}. For example, ~\citet{shen2022unifier} proposed UnifieR that optimizes a KL divergence loss between the predicted scores of two retrieval models besides the ranking task loss.

The aforementioned ensemble retrievers always show better performance than a single-model retriever in practice~\cite{nguyen2016ms, craswell2020overview}.
However, their component retrievers, no matter trained independently or jointly, are not handled specially according to their strengths and weaknesses in certain relevance patterns with respect to model architectures.
In contrast, built on the Mixture-of-Experts framework, our approach encourages the experts to compete with each other and specialize on suitable samples.

\subsection{Mixture-of-Experts Framework} \label{sec:related_work_MoE}
The Mixture-of-Experts (MoE) framework~\cite{jacobs1991adaptive} is one of the techniques for building ensemble models.
It builds upon the ``divide-and-conquer'' principle, in which the problem space is divided for several expert models, guided by one or a few gating networks.
According to how and when the gating network is involved in the dividing and combining procedures, MoE implementations could be classified into two groups~\cite{masoudnia2014mixture}, i.e., the mixture of implicitly localized experts (MILE) and the mixture of explicitly localized experts (MELE).
MILE divides the problem space implicitly using a tacit competitive process, such as using a special error function. 
MELE conducts explicit divisions by a pre-specified clustering method before the expert training starts, and each expert is then assigned to one of these sub-spaces.

The MoE framework has been widely used in various machine learning tasks~\cite{yuksel2012twenty,zhao2019recommending}, such as natural language processing and computer vision, showing great potential.
A major milestone in language modeling applications appears in~\cite{shazeer2017outrageously}, which encourages experts to disagree among themselves so that the experts can specialize on different problem sub-spaces.
Recently, MoE has also been applied in search tasks.
For example, ~\citet{nogueira2018new} built an MoE system for query reformulation, where the input query is given to multiple agents and each returns a corresponding reformulation. Then, the aggregator collects the ranking lists of all reformulated queries and gets the fusion results.
~\citet{dai2022mixture} proposed MoEBQA to alleviate the parameter competition between different types of questions. 
In addition, given the practical needs in product search, where the data comes from different domains, search scenarios, or product categories, the MoE framework is employed in various e-commerce platforms~\cite{zou2022automatic,sheng2021one}.  
For example, ~\citet{sheng2021one} claimed that different domains may share common user groups and items, and each domain has its unique data. Thus, it is difficult for a single model to capture the characteristics of various domains. To address this problem, they built a star topology adaptive recommender for multi-domain CTR prediction.

Similarly, due to various relevance matching patterns between queries and documents, we propose an MoE-based retrieval model that aims to exploit the expertise of different retrieval models to identify diverse relevance patterns.
As far as we know, this is the first work to utilize the MoE framework to first-stage retrieval.

\begin{figure*}[!t]
\setlength{\abovecaptionskip}{1pt}
\includegraphics[scale=0.37]{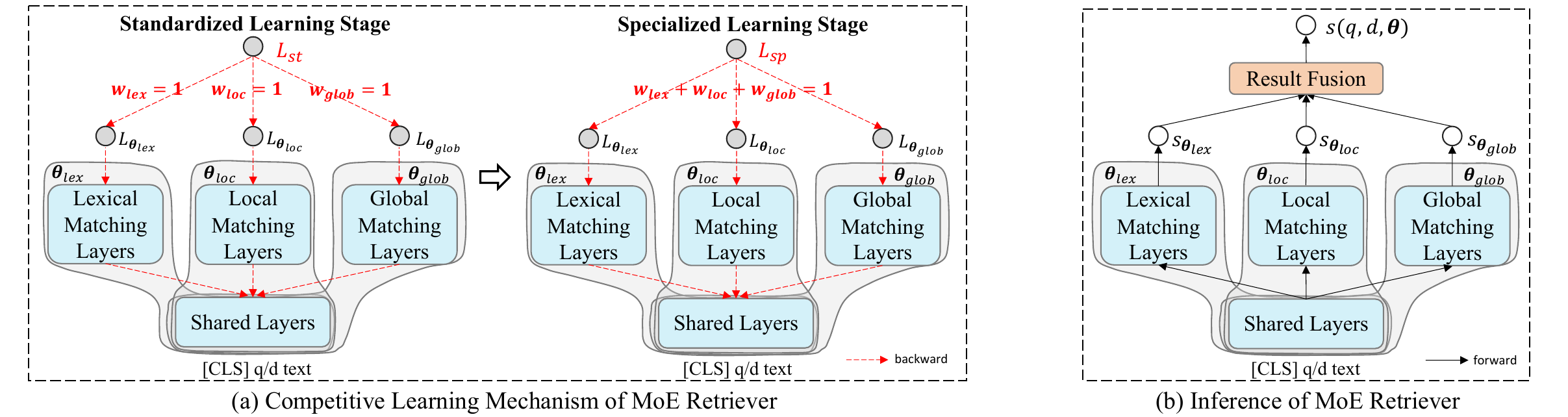}
\setlength{\belowcaptionskip}{-0.2cm}
\caption{The competitive learning mechanism (left) and inference (right) of the Mixture-of-Experts retriever. Since queries and documents share encoders, we only show one encoder for simplicity.}
\label{fig:model}
\end{figure*}

\section{Competitively Learn MoE Retriever}   \label{method}
In this section, we first describe the first-stage retrieval task. 
Then, we introduce the proposed MoE-based retrieval model, competitive learning mechanism, and inference procedure of CAME.

\subsection{Task Description}
Given a query $q$ and a collection $\smash{\mathcal{C}}$ with numerous documents, the first-stage retrieval aims to find as many potentially relevant documents as possible.
Formally, given a labeled training dataset $\smash{\mathcal{D}=\{(q_i, D_i^+)\}_{i=1}^{N}}$, where $q_i$ denotes a textual query, $\smash{D_i^+}$ is the set of relevant documents for $q_i$, and $N$ is the total number of queries. Let $\smash{d_i^+ \in D_i^+}$ be one of the relevant documents.
The essential task is to learn a model from $\mathcal{D}$ that gives high scores to relevant query-document pairs ($\smash{q_i}$, $\smash{d_i^+}$) and low scores to irrelevant ones. Then, all documents in the collection $\smash{\mathcal{C}}$ can be ranked according to the predicted scores. 
In the rest of this paper, we omit the subscript $i$ for a specific query if no confusion is caused.

\subsection{MoE Retriever}   \label{model}
To capture diverse relevance matching patterns, we include three representative retrievers in our MoE model, which are lexical, local, and global matching experts.
Due to the efficiency constraint in first-stage retrieval, we adopt the typical bi-encoder architecture~\cite{bromley1993signature} that encodes the query $q$ and the document $d$ separately, and then employs a simple interaction function to compute their relevance score based on the obtained representations:
\begin{equation}
\label{score}
s\left(q, d; \bm{\theta}\right)=\operatorname{f}\left(\operatorname{Enc_q}\left(q\right), \operatorname{Enc_d}\left(d\right)\right),
\end{equation}
where $\operatorname{Enc_q}$ and $\operatorname{Enc_d}$ denote the encoder modules for the query and document respectively, $\operatorname{f}$ is the interaction module, and the whole model is parameterized with $\bm{\theta}$. 
Following~\cite{xiong2020approximate, zhan2021optimizing}, the two encoders in \modelname{} have the same architecture and shared parameters. 
For each encoder, the bottom layers are shared for all the experts and the upper layers are private for them, as shown in Figure~\ref{fig:model} (a).
The relevance score of each query-document pair is calculated based on all the component expert predictions with a result fusion module, as shown in Figure~\ref{fig:model} (b).
We will elaborate on each part next.

\subsubsection{Shared Layers}
Existing studies~\cite{liu2019linguistic} have shown that the bottom layers of deep models tend to extract task-agnostic linguistic knowledge, so we use shared layers between the experts to capture basic textual features among all training samples. The learned common semantic and syntactic knowledge can act as the foundation to facilitate the upper layers to focus on different relevance patterns. 
Here, we leverage a stack of Transformer~\cite{vaswani2017attention} layers to produce contextualized token embeddings:
\begin{equation}
\begin{aligned}
&[\bm{c_{cls}^q};\bm{c_{1:m}^q}] = \operatorname{Trm}_{\scriptstyle share}\left(\left[\mathrm{CLS}; q\right]\right), \\ 
&[\bm{c_{cls}^d}; \bm{c_{1:n}^d}] = \operatorname{Trm}_{\scriptstyle share}\left(\left[\mathrm{CLS}; d\right]\right), \\
\end{aligned}
\end{equation}
where `[CLS]' is the special token prepended to the input text, and $m$ and $n$ denote the query and document length respectively.

\subsubsection{Mixture-of-Experts (MoE) Layers}
As shown in Figure~\ref{fig:model} (a), we employ three representative expert pathways at the upper layers, introduced as follows:

\textbf{Lexical Matching Expert.} To identify lexical matching, we incorporate a representative lexical retriever SPLADE~\cite{formal2021splade} as one of the expert pathways.
It maps each token vector output by the Transformer layer into a term-weighting distribution. The distribution is predicted based on the logits output by the Masked Language Model (MLM) layer~\cite{devlin2018bert}:
\begin{equation}
\begin{aligned}
&[\bm{h_{cls}^q}; \bm{h_{1:m}^q}] = \operatorname{MLM}(\operatorname{Trm}_{\scriptstyle lex}([\bm{c_{cls}^q}; \bm{c_1^q}; \cdots; \bm{c_m^q}])), \\
&[\bm{h_{cls}^d}; \bm{h_{1:n}^d}] = \operatorname{MLM}(\operatorname{Trm}_{\scriptstyle lex}([\bm{c_{cls}^d}; \bm{c_1^d}; \cdots; \bm{c_n^d}])). \\
\end{aligned}
\end{equation}
Then, the lexical representation for the query/document is obtained by aggregating the importance distributions over the whole input token sequence, and the relevance score is measured by dot-product:
\begin{equation}
\begin{aligned}
&\bm{e^q} = \max_{i=0,\cdots,m} \log (1+\operatorname{ReLU}(\bm{h_i^q})), \\
&\bm{e^d} = \max_{i=0,\cdots,n} \log (1+\operatorname{ReLU}(\bm{h_i^d})), \\
&s(q, d; \bm{\theta}_{\scriptstyle lex}) = \langle \bm{e^q} \cdot \bm{e^d} \rangle,
\end{aligned}
\end{equation}
where $\smash{\bm{h_{0}^q}}$ and $\smash{\bm{h_{0}^d}}$ correspond to $\smash{\bm{h_{cls}^q}}$ and $\smash{\bm{h_{cls}^d}}$ respectively. 

\textbf{Local Matching Expert.} To capture the local matching, we leverage the token-level query/document representations as ColBERT~\cite{khattab2020colbert} does in the second expert pathway. Specifically, it applies a linear layer on top of each token vector output by the final Transformer layer to compress the output dimension of local representations. Then, for every local query representation, it computes the maximum similarity (MaxSim) score across all the local document representations and sums these scores to obtain the final relevance:
\begin{equation}
\begin{aligned}
&[\bm{e_{cls}^q}; \bm{e_{1:m}^q}] = \operatorname{Trm}_{loc}([\bm{c_{cls}^q}; \bm{c_1^q}; \cdots; \bm{c_m^q}]) * \bm{W}, \\
&[\bm{e_{cls}^d}; \bm{e_{1:n}^d}] = \operatorname{Trm}_{loc}([\bm{c_{cls}^d}; \bm{c_1^d}; \cdots; \bm{c_n^d}]) * \bm{W}, \\
&s(q, d, \bm{\theta}_{\scriptstyle loc}) = \sum_{i=0}^{m} \max_{j=0,\cdots,n} \langle \bm{e_i^q} \cdot \bm{e_j^d} \rangle,
\end{aligned}
\end{equation}
where $\smash{\bm{e_{0}^q}}$ and $\smash{\bm{e_{0}^d}}$ correspond to $\smash{\bm{e_{cls}^q}}$ and $\smash{\bm{e_{cls}^d}}$ respectively, and $\bm{W}$ is the parameter of the shared linear layer.

\textbf{Global Matching Expert.} To produce the sequence-level representations and measure global matching, we design an expert pathway following DPR~\cite{karpukhin2020dense}. We extract the output vector of the `[CLS]' token as the global query/document representation, and define the relevance by the dot-product between the two vectors:  
\begin{equation}
\begin{aligned}
&\bm{e^q} = \operatorname{Pool}_{\scriptstyle cls}(\operatorname{Trm}_{glob} ([\bm{c_{cls}^q}; \bm{c_1^q}; \cdots; \bm{c_m^q}])), \\ 
&\bm{e^d} = \operatorname{Pool}_{\scriptstyle cls}(\operatorname{Trm}_{glob} ([\bm{c_{cls}^d}; \bm{c_1^d}; \cdots; \bm{c_n^d}])), \\
&s(q, d; \bm{\theta}_{\scriptstyle glob}) = \langle \bm{e^q} \cdot \bm{e^d} \rangle,
\end{aligned}
\end{equation}
where $\operatorname{Pool}_{\scriptstyle cls}$ means to take the output of the `[CLS]' token by the last Transformer layer of $\operatorname{Trm}_{glob}$.

\subsection{Competitive Learning Mechanism}  \label{learning}
In the real world, human experts are often trained through different stages of education, i.e., standardized training to gain general knowledge and specialized training to obtain expertise in a specific domain. Inspired by this process, our competitive learning mechanism has two similar stages, as shown in Figure~\ref{fig:model} (a). Remarkably, our specialized training is through competition. 

We first introduce the basic training objective of general retrievers as a preliminary.
They usually sample a set of negatives $\smash{D^-}$ from $\smash{\mathcal{C}}$ for a given relevant query-document pair $(q, \smash{d^+})$ in the training data. 
For $q$ and any $\smash{d \!\in\! \{d^+\} \!\cup\! D^-}$, the retriever parameterized by $\bm{\theta}$ computes the relevance score $s(q, d; \bm{\theta})$ with Eq.~(\ref{score}). 
Then, the probability distribution over the documents $\{\smash{d^+}\} \!\cup\! \smash{D^-}$ is defined as:
\begin{equation}
\label{probability}
P\left(d \!\mid\! q, d^{+}, D^-; \bm{\theta}\right)\!=\!\frac{\exp(s(q, d; \bm{\theta}))}{\sum_{d \in\left\{d^{+}\right\} \cup D^-} \exp (s(q, d; \bm{\theta}))}, \forall d \!\in\!\{d^{+}\} \!\cup\! D^-.
\end{equation}
Finally, a contrastive learning loss is used to optimize the retriever:
\begin{equation}
\label{basic_loss}
L_{\bm{\theta}} = -\log P\left(d^{+} \!\mid\! q, d^{+}, D^-; \bm{\theta}\right).
\end{equation}
As in~\cite{karpukhin2020dense}, in-batch negatives are generally used to facilitate the retriever learning. This objective is the basis for our two-stage competitive learning, which we will introduce next. 

\textbf{Standardized Learning Stage.}  
At the beginning, to encourage the experts to develop decent relevance matching capabilities and show potentials in certain types of samples, in the first few steps, we update the component experts by considering their loss equally important for every training sample in the standardized learning loss $\smash{L_{st}}$:
\begin{equation}
\label{warm_loss}
L_{st} = L_{\bm{\theta}_{\scriptstyle lex}} + L_{\bm{\theta}_{\scriptstyle loc}} + L_{\bm{\theta}_{\scriptstyle glob}}.
\end{equation}
At this stage, we sample the negative instances $\smash{D^-}$ in Eq.~(\ref{basic_loss}) from the top retrieved results of BM25~\cite{robertson2009probabilistic} excluding the relevant documents, denoted as $\smash{\mathcal{N}^{(bm25)}}$.

\textbf{Specialized Learning Stage.}
After the first stage, the experts show superiority on certain types of samples than the others. To enhance their expertise, in the second stage, we let them compete with each other on every training sample $(q, \smash{d^{+}}, \smash{D^-})$. The specialized learning loss $\smash{L_{sp}}$ is the weighted combination of the individual loss from each expert: 
\begin{equation}
\label{reinforcement_loss}
L_{sp}=\sum_{\text{\tiny \Ecommerce{}}} w_{\text{\tiny \Ecommerce{}}} \cdot L_{\bm{\theta}_{\text{\tiny \Ecommerce{}}}},
\end{equation}
where $\text{\scriptsize \Ecommerce{}} \in \{lex, loc, glob\}$, and $\smash{w_{\text{\tiny \Ecommerce{}}}}$ is the weight of the expert \scriptsize \Ecommerce{} \normalsize in the overall loss, computed according to its relative performance across all the experts on this sample:
\begin{equation}
\label{weight}
w_{\text{\tiny \Ecommerce{}}} = \operatorname{Softmax}_{\text{\tiny \Ecommerce{}}}(\frac{1}{Rank\left(d^{+} \!\mid\! q, d^{+}, D^-; \bm{\theta}_{\text{\tiny \Ecommerce{}}}\right)}; \tau).
\end{equation}
In Eq.~(\ref{weight}), $Rank\left(d^{+} \!\mid\! q, d^{+}, D^-; \bm{\theta}_{\text{\tiny \Ecommerce{}}}\right)$ denotes the rank of $d^+$ in expert \scriptsize \Ecommerce \normalsize's predictions, which represents how effectively this expert can perform on the training instance $(q,\smash{d^+},\smash{D^-})$; 
$\tau$ is the temperature to control the degree of soft competition, where a smaller $\tau$ means a more intense competition; and the softmax function is applied on all the experts so that $\sum_{\text{\tiny \Ecommerce{}}} w_{\text{\tiny \Ecommerce}} = 1$ for each training sample and the experts compete for the update weights. 

Similar to~\cite{xiong2020approximate,zhan2021optimizing}, we train \modelname{} according to Eq.~(\ref{reinforcement_loss}) first on BM25 negatives until convergence and then on hard negatives to enhance its performance. 
We collect hard negatives from \modelname{} learned on BM25 negatives.
Specifically, for each query, we first obtain the top retrieval results of every expert \scriptsize \Ecommerce{} \normalsize excluding the relevant ones, denoted as $\mathcal{N}^{(\text{\tiny \Ecommerce{}})}$. Then, we sample negatives $D^-$ from $\mathcal{N}^{({hard})} = \mathcal{N}^{({lex})} \cup \mathcal{N}^{(loc)} \cup \mathcal{N}^{(glob)}$ and further train \modelname{} with Eq.~(\ref{reinforcement_loss}).

\subsection{MoE Inference}
During inference time (as shown in Figure~\ref{fig:model} (b)), each expert that specializes on a specific relevance pattern can judge from its perspective to determine the relevance scores of documents. Since which expert is more trustworthy for an unseen query-document pair is unknown, we simply consider them equally authoritative. 
Specifically, for a test query $q$, to obtain its top-$K$ retrieval results of \modelname{}, we first get the top-$K$ results of every component expert \scriptsize \Ecommerce{} \normalsize from the whole collection $\smash{\mathcal{C}}$ according to the predicted score $\smash{s(q, d, \bm{\theta}_{\text{\tiny \Ecommerce{}}})}$, then we sum up the document scores from each expert to determine the final top-$K$ results:
\begin{equation}
\label{ensemble_score}
s(q, d; \bm{\theta})=s(q, d; \bm{\theta}_{\scriptstyle lex}) + s(q, d; \bm{\theta}_{\scriptstyle loc}) + s(q, d; \bm{\theta}_{\scriptstyle glob}).
\end{equation}
The score of a document that is beyond top-$K$ results of an expert is set to the  score of the $K$th document this expert has on $q$.
This simple method performs similarly or better compared to other fusion methods, e.g., combining expert results based on the summation of reciprocal ranks and learning a linear combination of the expert scores (see Table~\ref{tab:fusion}).
We leave the study of more complex fusion methods in the future, e.g., learning to predict the probability of  each relevance pattern given the query and document, and assign the authority weights to each expert for the fusion.

\subsection{Discussions}
\noindent \textbf{Competitive Learning versus Collaborative Learning.}
Previous ensemble methods, such as DrBoost~\cite{lewis2021boosted} and UnifieR~\cite{shen2022unifier}, usually adopt collaborative learning mechanisms which directly optimize the overall score linearly combined from the component retrievers~\cite{lewis2021boosted} or optimize one component towards the other based on a KL divergence loss~\cite{shen2022unifier}. 
In contrast, the competitive learning we propose has the advantage of better fitting each expert to its suitable samples since:
(1) the learning of each expert is based on its own error;
(2) the update weight for each expert is proportional to how it performs compared to other experts. 
Given a training sample, when one of the experts produces less error than others, it indicates that this sample is likely to match the expert's expertise, thus its responsibility for the sample will be increased, and vice versa.
Accordingly, each expert is strengthened with suitable samples and avoids forcing itself to learn from the unsuitable ones, which further enhances its expertise in identifying corresponding relevance patterns.

\vspace*{0.5mm}
\noindent \textbf{Implicitly versus Explicitly MoE.}
With the competitive learning objective, i.e., Eq.~(\ref{reinforcement_loss}\&\ref{weight}), \modelname{} implicitly divides the problem space into a number of sub-spaces, and the experts are trained to specialize in different sub-spaces.
According to the taxonomy of MoE models (see details in Section~\ref{sec:related_work_MoE}), \modelname{} obviously falls into the category of the mixture of implicitly localized experts (MILE)~\cite{masoudnia2014mixture}. 
Since the problem sub-spaces could have overlaps (multiple relevance patterns could exist for a sample), it is more reasonable to implicitly guide the experts towards suitable samples according to their own predictions compared to training a mixture of explicitly localized experts (MELE) with pre-specified explicit divisions of samples.

\vspace*{0.5mm}
\noindent \textbf{Runtime Overhead.}
During runtime, for ensemble models like \modelname{}, since the component retrievers can perform retrieval in parallel, the query response time would not be a major concern.
Also, the common practice of using approximate nearest neighbor search algorithms, such as product quantization~\cite{jegou2010product}, can significantly decrease the time and memory costs of combining multiple models.
Moreover, for the local matching expert that has the highest runtime overhead, i.e., ColBERT, existing methods of improving it, e.g., token pruning~\cite{tonellotto2021query, lassance2022learned}, can be easily applied to  \modelname{} to further reduce the overhead. 

\vspace*{-3mm}
\section{Experimental Settings} 
This section presents the experimental settings, including datasets, baselines, evaluation metrics, and implementation details.

\vspace*{-2mm}
\subsection{Datasets}
We conduct experiments on three retrieval benchmarks: 
\begin{itemize}[leftmargin=*,topsep=0pt,parsep=0pt]
\item \textbf{MS MARCO~\cite{nguyen2016ms}:} MS MARCO passage ranking task is introduced by Microsoft. It focuses on ranking passages from a collection with over 8.8M passages.
It has about 503k training queries and about 7k queries in the dev/test set for evaluation. Since the test set is not available, we report the results on the dev set.
\item \textbf{TREC DL~\cite{craswell2020overview}:} TREC 2019 Deep Learning Track has the same training and dev set as MS MARCO, but replaces the test set with a novel set produced by TREC. It contains 43 test queries for the passage ranking task.
\item \textbf{NQ~\cite{kwiatkowski2019natural}:} Natural Questions collects real questions from Google's search logs. Each query is paired with an answer span and golden passages in Wikipedia pages. There are more than 21 million passages in the corpus. It also has over 58.8k, 8.7k, and 3.6k queries in training, dev, and test sets respectively. In our experiments, we use the version of NQ created by~\citet{karpukhin2020dense}.
\end{itemize}

\vspace{-3mm}
\subsection{Baselines}  \label{sec:baseline}
We consider two types of baselines for performance comparison, including representative single-model and ensemble retrievers.

\noindent \textbf{Representative single-model retrievers:}
\begin{itemize}[leftmargin=*,topsep=0pt,parsep=0pt]
\item \textbf{SPLADE~\cite{formal2021splade}} is the exemplar of lexical retrievers. It projects the input text to a $|V_{BERT}|=30522$ dimensional sparse representation. We set $\lambda=0.01$ for the FLOPS regularizer on NQ. 
\item \textbf{ColBERT~\cite{khattab2020colbert}} is the exemplar of local retrieval models, which measures relevance by soft matching all query and document tokens’ contextualized vectors. To run ColBERT on TREC DL and NQ, we set the output dimension as 128. 
\item \textbf{COIL~\cite{gao2021coil}} is a variant of ColBERT, which only calculates similarities between exact matched terms for queries and documents in the MaxSim operator. We use the output dimension of local representations with 32. 
\item \textbf{DPR~\cite{karpukhin2020dense}} is the basic global retrieval model with a BERT-base bi-encoder architecture.
\item \textbf{RocketQA~\cite{qu2020rocketqa}} is one of the state-of-the-art single-model retrievers, which utilizes three training strategies to enhance the basic global retriever, namely cross-batch negatives, denoised hard negatives, and data augmentation. 
\item \textbf{AR2~\cite{zhang2021adversarial}} is the state-of-the-art single-model retriever, which consists of a global retriever and a cross-encoder ranker jointly optimized with a minimax adversarial objective. 
\end{itemize}

\noindent \textbf{Ensemble retrievers:}
\begin{itemize}[leftmargin=*,topsep=0pt,parsep=0pt]
\item \textbf{ME-HYBRID~\cite{luan2021sparse}} first trains a local retriever and a lexical retriever independently, and then linearly combines them by fusing their predicted scores. 
\item \textbf{SPAR~\cite{chen2021salient}} trains a global retriever (namely $\bm{\Lambda}$) to imitate a lexical one, and combines $\bm{\Lambda}$ with another global retriever to enhance the lexical matching capacity. Here we use the weighted concatenation variant for implementation, where the two component retrievers are trained independently. 
\item \textbf{CLEAR~\cite{gao2021complement}} trains a global retriever with the mistakes made by a lexical retriever to combine the capacities of both models collaboratively. 
\item \textbf{DrBoost~\cite{lewis2021boosted}} is an ensemble retriever inspired by boosting technique. Each component retriever is learned sequentially and specialized by focusing only on the retrieval mistakes made by the current ensemble. The final representation is the concatenation of the output vectors from five component models.
\item \textbf{DSR~\cite{lin2021densifying}} densifies the representations from a lexical retrieval model and combines them with a global retriever for joint training. We use the variant of ``DSR-SPLADE + Dense-[CLS]''. 
\item \textbf{UnifieR~\cite{shen2022unifier}} is a latest and state-of-the-art method that unifies a global and a lexical retriever in one model and aligns each component retriever with the other based on their listwise agreements on the predictions.  
\end{itemize}

\renewcommand{\arraystretch}{1.3}
\begin{table*}[!t]
\setlength\tabcolsep{8pt}
\setlength{\abovecaptionskip}{0pt}
  \centering
  \fontsize{9}{9}\selectfont
    \caption{Overall first-stage retrieval performance. Model types are shown in parentheses ($\bm{lex}$: lexical retriever, $\bm{loc}$: local retriever, $\bm{glob}$: global retriever, $\bm{indep}$: independent learning, $\bm{joint}$: joint learning). 
    Bold and \underline{underline} indicate the best overall and baseline performance respectively. The statistically significant (p < 0.05) improvements of \modelname{} over the best baselines RocketQA, AR2, SPAR, and DSR are marked with $\dag$, $\ddag$, $\diamond$, and $\ast$ respectively.} 
  \label{tab:performance_comparison}
  \begin{threeparttable}
    \begin{tabular}{lllllll}
    \toprule
    \multirow{2}*{Method}
    &\multicolumn{2}{c}{MS MARCO Dev}&\multicolumn{2}{c}{TREC DL Test}&\multicolumn{2}{c}{NQ Test}\cr
    \cmidrule(lr){2-3} \cmidrule(lr){4-5} \cmidrule(lr){6-7} 
    &R@1000 &MRR@10 &R@1000 &NDCG@10 &Top-20 &Top-100\cr
    \midrule
    \textit{Single-model Retriever} & & & & &  &  \cr
    $\bm{(lex)}$ BM25 &85.3 &18.4 &74.5 &50.6 &59.1 &73.7 \cr
    $\bm{(lex)}$ SPLADE~\cite{formal2021splade} &96.5 &34.0 &\underline{85.1} &68.4  &77.4 &84.8 \cr
    $\bm{(loc)}$ ColBERT~\cite{khattab2020colbert}  &96.8 &36.0  &72.7 &68.7 &78.8 &84.6  \cr
    $\bm{(loc)}$ COIL~\cite{gao2021coil}  &96.3 &35.5  &84.8 &\underline{71.4}   &77.9 &84.5 \cr
    $\bm{(glob)}$ DPR~\cite{karpukhin2020dense} &94.1 &31.6 &70.4 &61.6  &78.4 &85.4  \cr
    $\bm{(glob)}$ RocketQA~\cite{qu2020rocketqa} &97.9 &37.0  &79.3 &71.3  &82.7 &88.5  \cr
    $\bm{(glob)}$ AR2~\cite{zhang2021adversarial} &\underline{98.6} &\underline{39.5}  &82.0 &70.0  &\underline{86.0} &\underline{90.1}  \cr
    \midrule
    \textit{Ensemble Retriever} & & & &  & &   \cr
    $\bm{(indep)}$ ME-HYBRID~\cite{luan2021sparse} &97.1 &34.3  &81.7 &70.6   &82.6 &88.6 \cr
    $\bm{(indep)}$ SPAR~\cite{chen2021salient} &\underline{98.5} &38.6 &82.8 &71.7  &83.0 &88.8 \cr
    $\bm{(joint)}$ CLEAR~\cite{gao2021complement}  &96.9 &33.8  &81.2 &69.9  &82.3 &88.5   \cr
    $\bm{(joint)}$ DrBoost~\cite{lewis2021boosted} &- &34.4 &- &-  &80.9 &87.6 \cr
    $\bm{(joint)}$ DSR~\cite{lin2021densifying} &96.7 &35.8  &\underline{84.1} &69.4  &\underline{83.2} &\underline{88.9} \cr
    $\bm{(joint)}$ UnifieR~\cite{shen2022unifier}  &98.4 &\underline{40.7}  &- &\underline{73.8}  &- &-   \cr
    \midrule
    $\bm{(joint)}$ \modelname{} &$\bm{98.8}^{\dag \ast}$ &$\bm{41.3}^{\dag \ddag \diamond \ast}$ &$\bm{86.6}^{\dag \ddag \diamond \ast}$ &$\bm{74.5}^{\dag \ddag \diamond \ast}$ &$\bm{87.5}^{\dag \ddag \diamond \ast}$ &$\bm{91.4}^{\dag \ddag \diamond \ast}$   \cr 
    \bottomrule
    \end{tabular}
    \end{threeparttable}
\vspace{-2mm}
\end{table*}

\subsection{Evaluation Metrics}
We use the official metrics of the three benchmarks.
For retrieval tasks (i.e., MS MARCO and TREC DL), the retrieval performance of top 1000 passages is compared.
We report the Recall at 1000 (R@1000) and Mean Reciprocal Rank at 10 (MRR@10) for MS MARCO, and R@1000 and Normalized Discounted Cumulative Gain at 10 (NDCG@10) for TREC DL. 
For open-domain question answering tasks (i.e., NQ), the proportion of top-$n$ retrieved passages that contain the answers (Top-$n$) is compared~\cite{karpukhin2020dense}, and we report Top-20 and Top-100.
Statistically significant differences are measured by two-tailed t-test.

\subsection{Implementation Details} 
As in UnifieR~\cite{shen2022unifier}, for the Transformer and MLM layers in \modelname{}, we initialize the parameters with the coCondenser~\cite{gao2021unsupervised} checkpoint released by Gao et al.\footnote{\url{https://github.com/luyug/Condenser}}.
Unless otherwise specified, the number of Transformer layers in the shared layers and MoE layers are set to 10 and 2 respectively.
We adopt the popular Transformers library\footnote{\url{https://github.com/huggingface/transformers}} for implementations. 
For MS MARCO and TREC DL, we truncate the input query and passage to a maximum of 32 tokens and 128 tokens respectively. We train \modelname{} with the official BM25 top 1000 negatives for 5 epochs (including 1 epoch for standardized learning and 4 epochs for specialized learning), and then mine hard negatives from the top 200 retrieval results to train 3 more epochs. We use a learning rate of 5e-6 and a batch size of 64. The temperature $\tau$ is tuned on a held-out training subset and is set to 0.5.
For NQ, we set the query and passage length to 32 and 156 respectively. We first train \modelname{} with the official BM25 top 100 negatives for 15 epochs (including 3 epochs for standardized learning and 12 epochs for specialized learning), and then mine hard negatives from the top 100 retrieval results for 10-epoch further training. We use a learning rate of 1e-5 and a batch size of 256. The temperature $\tau$ is tuned on the dev set and is set to 2.0.
For all these three datasets, each positive example is paired with 7 negatives for training. We control the sparsity of the lexical representations in the lexical matching expert with $\lambda=0.01$ for the FLOPS regularizer~\cite{formal2021splade}, and the output dimension of local representations in the local matching expert is set to 128.
We use the AdamW optimizer with $\beta_1 = 0.9$, $\beta_2 = 0.999$, $\epsilon = 10^{-8}$, and set the coefficient to control linear learning rate decay to 0.1. 
We run all experiments on Nvidia Tesla V100-32GB GPUs.

For all the baseline models, we use the official code or checkpoint to reproduce the results, except for CLEAR, DrBoost, and UnifieR whose code is not available, so we report the evaluation results from their paper directly\footnote{We report the performance of CLEAR, DrBoost, and UnifieR from their original papers and we could not conduct significance test against them since their code is not released and their result lists are not available.}.

\section{Results and Discussion} \label{results}
In this section, we present the experimental results and conduct thorough analysis of \modelname{} to  answer the following research questions:
\begin{itemize}[leftmargin=*,topsep=0pt,parsep=0pt]
\item \textbf{RQ1:} How does \modelname{} perform compared to the baseline methods, including state-of-the-art single-model retrievers and ensemble retrievers?
\item \textbf{RQ2:} How does each of the two stages in the training pipeline affect the retrieval performance?
\item \textbf{RQ3:} How does each component in \modelname{} contribute to the retrieval performance?
\item \textbf{RQ4:} How do the key hyper-parameters in \modelname{} affect the retrieval effectiveness?
\end{itemize}

\subsection{Main Evaluation}
To answer \textbf{RQ1}, we compare the retrieval performance of \modelname{} with all the baselines described in Section~\ref{sec:baseline} and record their evaluation results in Table~\ref{tab:performance_comparison}.

\textbf{Comparisons with Single-model Retrievers.}  
The top block of Table~\ref{tab:performance_comparison} shows the performance of the representative baseline retrieval models.
From the results on three datasets, we find that:
(1) Neural retrieval models built upon pre-trained language models perform significantly better than term-based retrieval models, i.e., BM25, except for R@1000 on TREC DL.
(2) Among all the neural retrievers, RocketQA and AR2 have the best performance on MS MARCO and NQ probably due to their more sophisticated training strategies, such as the data augmentation (in RocketQA) and knowledge distillation (in AR2).
(3) For the other methods without complex training strategies, we find that the models that achieve the best performance on each dataset are of different types. Local retriever ColBERT has the best performance on MS MARCO while SPLADE and COIL perform the best on TREC DL. SPLADE is a lexical retriever and COIL is the variant of ColBERT that only keeps the similarities between the exact matched query and document terms.  Global retriever DPR has the overall best performance on NQ. This observation also supports our claim that various relevance matching patterns exist. 
(4) By leveraging multi-types of retrievers, \modelname{} outperforms all the single-model baselines by a large margin in terms of almost all the metrics, showing the superiority of capturing various relevance patterns. Note that \modelname{} should achieve even better performance if using the knowledge distillation and data augmentation techniques proposed in RocketQA and AR2.  


\textbf{Comparisons with Ensemble Retrievers.} 
The performance of the ensemble retrievers are presented in the bottom block of Table~\ref{tab:performance_comparison}.
We have the following observations:
(1) An ensemble retriever that consists of only a single type of model underperforms ensemble retrievers that have multi-types of component models. This can be seen from that DrBoost (only using global retrievers) performs worse than the others. 
(2) Among the methods that combine various types of retrievers (i.e., ensemble models except DrBoost), DSR and UnifieR perform the best. The superiority of DSR and UnifieR should mostly come from the elaborate joint learning strategies, which shows the importance of the learning mechanism.
(3) Compared to the existing ensemble retrievers, \modelname{} achieves significantly better performance on the three datasets in terms of almost all the metrics. For example, its performance improvements over UnifieR are 1.5\% and 0.9\% on MRR@10 and NDCG@10 for MS MARCO and TREC DL respectively, and the improvements over DSR are 5.1\% and 2.8\% regarding Top-20 and Top-100 for NQ.
These performance gains further confirm \modelname{}'s advantages from employing multi-types of retrievers to capture various relevance patterns and the competitive learning mechanism that facilitates the component expert training. 


\subsection{Ablation Study of Learning Mechanism}
To answer \textbf{RQ2}, we conduct ablation studies on each stage of the learning pipeline. 
We only report the results on MS MARCO and NQ in Table~\ref{tab:learning_strategy} due to the space limit, and TREC DL has similar trends. Since R@1000 on MS MARCO is already near 100\% and slightly differs between the variants, we also report R@100 to show the differences.

\renewcommand{\arraystretch}{1.1}
\setlength\tabcolsep{2.5pt}
\begin{table}[!t]
\setlength{\abovecaptionskip}{-0pt}
  \centering
  \fontsize{9}{9}\selectfont
  \caption{Ablation study of the competitive learning mechanism. Statistically significant differences (p < 0.05) over \modelname{} and the variant without hard negatives are marked with $\dag$ and $\ddag$ respectively.}
  \label{tab:learning_strategy}
  \begin{threeparttable}
  \begin{tabular}{llllll}
    \toprule
     \multirow{2}*{Method}
     &\multicolumn{3}{c}{MS MARCO}&\multicolumn{2}{c}{NQ} \cr
     \cmidrule(lr){2-4} \cmidrule(lr){5-6}
     &R@1000 &R@100 &MRR@10 &Top-20 & Top-100  \cr 
    \midrule
    \modelname{}  &98.8 &92.5 &41.3 &87.5 &91.4  \cr 
     - hard negatives &98.4  &$91.7^{\dag}$ &$39.0^{\dag}$ &$83.7^{\dag}$ &$89.1^{\dag}$   \cr 
     \quad - specialized &98.2 &$90.6^{\dag \ddag}$ &$36.5^{\dag \ddag}$  &$82.0^{\dag \ddag}$ &$87.8^{\dag \ddag}$   \cr 
     \quad - standardized &98.3 &$91.2^{\dag}$ &$37.9^{\dag \ddag}$  &$82.5^{\dag \ddag}$ &$88.1^{\dag \ddag}$   \cr 
    \bottomrule
  \end{tabular}
\end{threeparttable}
\end{table}

We first remove the training based on hard negatives (i.e., w/o hard negatives) and see that the performance decreases by a large margin, especially on the top results (e.g., MRR@10 on MS MARCO and Top-20 on NQ). This observation is consistent with previous studies, showing that hard negatives could greatly enhance neural retrieval models~\cite{qu2020rocketqa}.
It should be noted that, even without using hard negatives for further learning, \modelname{} still outperforms most baselines (see  Table~\ref{tab:performance_comparison}), and UnifieR~\cite{shen2022unifier}, the best baseline which is reported to have 98.0 and 38.3 for R@1000 and MRR@10  under the same setting of using BM25 negatives only~\cite{shen2022unifier}.
It demonstrates the superiority of competitive learning in \modelname{}.

Based on the above variant of \modelname{} without hard negatives, we run another two ablation studies by: (1) removing the specialized learning stage and training \modelname{} until convergence with Eq.~(\ref{warm_loss}) only (i.e., w/o specialized); (2) removing the standardized learning stage and training \modelname{} with Eq.~(\ref{reinforcement_loss}) from the beginning (i.e., w/o standardized).
The first ablation shows that training without specialized learning leads to a substantial drop in retrieval performance, indicating that it is important to encourage component experts to compete with each other and focus on the samples they are skilled at.
The second one shows that training only with the specialized learning stage would also lower the performance. 
Without the standardized learning process, the component experts are not readily prepared to specialize on certain relevance patterns. Then, the update weights calculated with Eq.~(\ref{weight}) cannot reflect accurately how each training sample corresponds to the relevance matching patterns, which further makes it difficult to facilitate the experts to develop their own advantages based on suitable samples.

\subsection{Study of Model Component Choices}
To answer \textbf{RQ3}, we conduct ablation studies and compare with other alternatives for each component in \modelname{}. 
Since NQ and TREC DL lead to similar conclusions, we only report the performance on MS MARCO. Note that we only show the model performance based on BM25 negatives training due to the cost of collecting  hard negatives.

\renewcommand{\arraystretch}{1.1}
\setlength\tabcolsep{3pt}
\begin{table}[!t]
\setlength{\abovecaptionskip}{0pt}
  \centering
  \fontsize{9}{9}\selectfont
  \caption{Ablation study of the encoder module in \modelname{}. R@1000 and MRR@10 of the results retrieved by each expert and after fusion on MS MARCO are reported.} 
  \label{tab:model}
  \begin{threeparttable}
  \begin{tabular}{lcccccccc}
  \toprule
      \multirow{2}*{Method}
    &\multicolumn{2}{c}{Fusion}&\multicolumn{2}{c}{Lexical}&\multicolumn{2}{c}{Local}&\multicolumn{2}{c}{Global}\cr
    \cmidrule(lr){2-3} \cmidrule(lr){4-5} \cmidrule(lr){6-7} \cmidrule(lr){8-9} 
    &R &MRR &R &MRR &R &MRR &R &MRR\cr
    \midrule
    \modelname{}  &98.4 &39.0 &97.6 &36.1 &96.7 &38.8 &97.7 &36.8 \cr 
    \midrule
     - shared layers &98.3 &38.1 &97.4 &35.1 &96.0 &38.0 &97.7 &36.0    \cr 
     \midrule
     - lexical &98.1 &38.1 &- &- &96.5 &38.5 &97.8 &36.2    \cr 
     - local &98.2 &37.8 &97.4 &36.1 &- &- &97.7 &36.7    \cr 
     - global  &98.1 &38.0 &97.5 &35.9 &96.5 &38.7 &- &-    \cr
     \midrule
     - local\&global &97.3 &35.1 &97.3 &35.1 &- &- &- &- \cr
     - lexical\&global &96.0 &37.9 &- &- &96.0 &37.9 &- &- \cr
     - lexical\&local &97.7 &35.9 &- &- &- &- &97.7 &35.9 \cr
    \bottomrule
  \end{tabular}
\end{threeparttable}
\end{table}

\begin{figure*}[!t]
\setlength{\abovecaptionskip}{-0cm}
\includegraphics[scale=0.45]{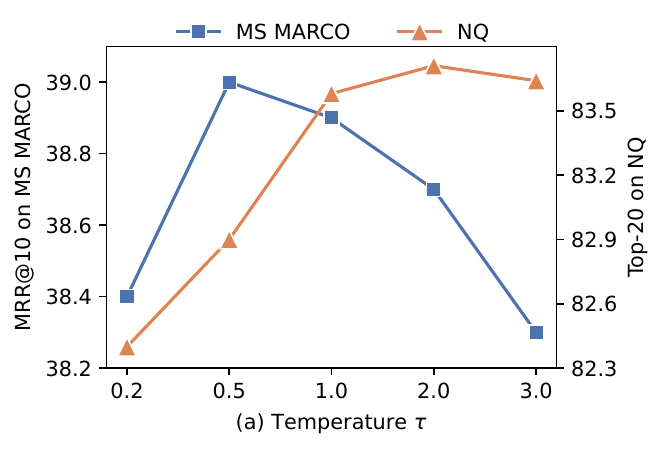}
\hspace{1.5mm}
\includegraphics[scale=0.45]{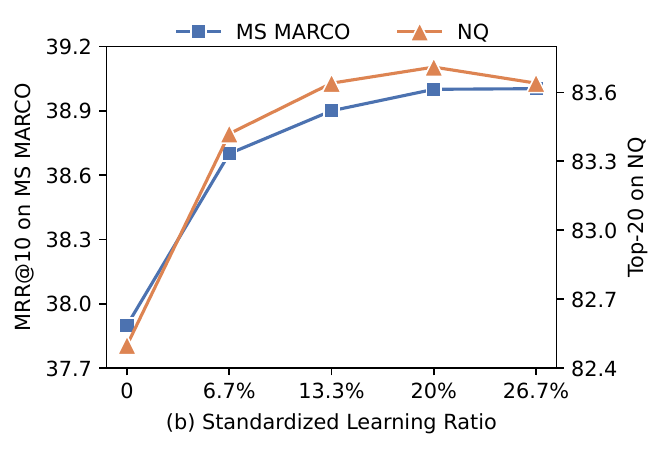}
\hspace{1.5mm}
\includegraphics[scale=0.45]{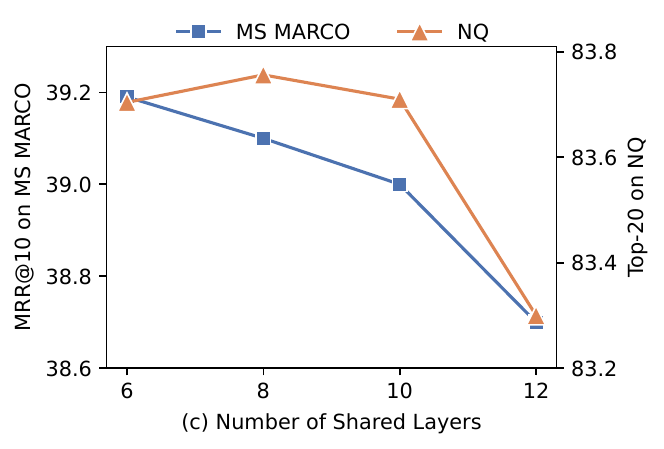}
\caption{The performance of \modelname{} with different hyper-parameter settings on MS MARCO and NQ.}
\label{fig:parameter}
\end{figure*}

\textbf{Ablation Study of Encoder Module.}
For the ablation study of the encoder module, we: (1) remove the shared layers, that is, each expert has private 12 Transformer encoding layers; (2) remove one expert in the MoE layers separately; (3) remove two experts in the MoE layers simultaneously, that is, the remaining one expert is trained independently. We report the performance of overall and each component expert under every ablation setting in Table~\ref{tab:model}.
By removing the shared layers, we find that MRR@10 of the fusion result decreases significantly, and the performance of every component expert is close to that of their own independent training. 
This shows that the shared bottom layers in \modelname{} act as multi-task learning that can facilitate the semantic and syntactic knowledge to be better learned for relevance estimation.
When removing any expert in \modelname{}, the fusion performance degrades dramatically. 
Meanwhile, using two experts in MoE layers could achieve better performance than using either of them independently, e.g., the lexical expert and global expert achieve 35.1 and 35.9 on MRR@10 independently (in the bottom block), while using both of them (i.e., the variant of w/o local expert) could achieve 37.8. 
Furthermore, the rank-biased overlap (RBO)~\cite{webber2010similarity} between top 1000 results retrieved by any two independently trained experts is 42\%-48\%.
It shows that documents may have multiple relevance patterns and the retrieval results  of different experts could complement each other.

\textbf{Study of Various Fusion Methods.}
To study the result fusion module, we compare  several variants with \modelname{} to verify our choice (see the top block of Table~\ref{tab:fusion}): (1) using summation/maximum of min–max normalized scores (\textit{NormSum/NormMax}) instead of the summation of raw scores (\textit{Sum}) in Eq.~(\ref{ensemble_score}); (2) using reciprocal ranks (\textit{SumRR/MaxRR}) instead of scores for fusion; (3) conducting learnable linear combination of model scores (\textit{LinearLayer}).
We can see that \textit{NormSum} and \textit{LinearLayer} do not outperform the simple score summation, i.e., \textit{Sum}, in \modelname{}.
We also find that the fusion of reciprocal ranks has worse MRR@10 than others, probably due to their faster decay than scores.
Also, maximum always yields worse results than summation, probably because multiple relevance patterns co-exist in some documents and the judgment from each expert is important to evaluate the overall relevance.

\textbf{Study of Alternative Expert Learning Methods.}
To show the advantages of competitive learning, we compare it with alternative options of training an ensemble model with SPLADE, ColBERT, and DPR. We: (1) learn the three models independently on the entire training data that contains both suitable and unsuitable samples for each model (\textit{All-Independent}) and fuse their results with score summation, reciprocal rank summation, and a learned linear combination layer; (2) learn the three independent models on their suitable samples (\textit{Suit-Independent}), determined by whether the model has the best MRR@10 for the sample among all experts in \modelname{}. 
As shown in the bottom block of Table~\ref{tab:fusion}, the fusion of reciprocal ranks performs better than scores, which is different from our observations on fusing expert results in CAME. The possible reason is that the scores obtained from the independently learned models are less comparable than those achieved by jointly learned experts in CAME. Both of them are significantly worse than \modelname{} and the fusion variants based on competitive learning.
In addition, when fusing the model results with linear combination, models trained only on suitable samples (\textit{Suit-Independent}) have better ensemble performance than models trained on the whole data (\textit{All-Independent}), which is consistent with the performance gains produced by the specialized learning mechanism in CAME.

\renewcommand{\arraystretch}{1.1}
\setlength\tabcolsep{3pt}
\begin{table}[!t]
\setlength{\abovecaptionskip}{2pt}
  \centering
  \fontsize{9}{9}\selectfont
  \caption{Investigation of variou options to learn experts and fuse their results on MS MARCO. $\dag$ marks statistically significant differences (p\! <\! 0.05) with \modelname{} (Competitive\! +\! Sum).}
  \label{tab:fusion}
  \begin{threeparttable}
  \begin{tabular}{ccccc}
  \toprule
      {Learning Method} &{Fusion Method} &R@1000 &R@100 &MRR@10 \cr
    \midrule
    $\bm{Competitive}$  &$\bm{Sum}$ &98.4 &91.7 &39.0 \cr 
    \midrule
    \midrule
    Competitive  &NormSum &98.4 &91.7  &38.9 \cr
    Competitive  &NormMax &98.2 &$90.8^{\dag}$ &$37.9^{\dag}$ \cr 
    Competitive  &SumRR &98.5 &91.7 &38.4 \cr
    Competitive  &MaxRR &98.4 &91.3 &$38.1^{\dag}$ \cr
    Competitive  &LinearLayer  &98.4 &91.8  &39.2 \cr
    \midrule
    \midrule
    All-Independent  &Sum &$97.5^{\dag}$ &$88.4^{\dag}$  &$35.1^{\dag}$ \cr
    All-Independent  &SumRR &98.1 &$90.4^{\dag}$ &$37.2^{\dag}$ \cr
    All-Independent  &LinearLayer  &98.2 &$90.8^{\dag}$  &$37.8^{\dag}$ \cr
    Suit-Independent  &LinearLayer  &98.1 &$91.0^{\dag}$  &$38.1^{\dag}$ \cr
    \bottomrule
  \end{tabular}
\end{threeparttable}
\end{table}

\subsection{Hyper-parameter Impact}
To answer \textbf{RQ4}, we evaluate \modelname{} with different hyper-parameter settings to investigate their impact on retrieval results.
Again, we conduct analysis on the models trained on BM25 negatives due to the cost of collecting hard negatives.

\textbf{Impact of the Temperature.}
As shown in Figure~\ref{fig:parameter} (a), we examine the effect of temperature $\tau$ in Eq.~(\ref{weight}) by varying it from 0.2 to 3.0. The results indicate that $\tau$ has a great impact on retrieval performance, with different optimal values for the two datasets (0.5 for MS MARCO and 2.0 for NQ). MS MARCO has shorter documents than NQ on average, which could lead to that the performance of the local and global matching experts may not differ much on MS MARCO, thus it may require a smaller $\tau$ to make the competition between the experts more intensely to enhance their expertise.

\textbf{Impact of the Standardized Learning Ratio.}
Figure~\ref{fig:parameter} (b) shows how MRR@10 changes with respect to the standardized learning ratio, where the ratio means the proportion of the standardized learning steps in the training based on BM25 negatives.
For both datasets, the performance reaches the top with 20\% standardized learning steps. The performance is relatively stable when the ratio is between 13.3\% and 26.7\%, while decreasing dramatically at smaller values.
We speculate that the component experts cannot develop their expertise if they have not seen enough training samples, and once the standardized learning is enough, the ratio has a small effect on the retrieval effectiveness.

\textbf{Impact of the Number of Shared Layers.}
As shown in Figure~\ref{fig:parameter} (c), we investigate the model performance by varying the number of shared layers from 6 to 12.
On both MS MARCO and NQ, the performance decreases dramatically when the number of shared layers is set to 12. This indicates that the component experts cannot develop their expertise when their private parameters are too few.
In addition, we observe that with fewer shared layers, the performance becomes slightly better on MS MARCO, while 8 shared layers are better than 6 on NQ. The reason may be that fewer private layers are sufficient to learn effective experts for NQ due to its smaller size. 
Overall, the performance varies slightly from 6 to 10 shared layers  (i.e., within 0.2\%). Therefore, considering the balance between the effectiveness and efficiency of encoding, we set the number of shared layers as 10 in \modelname{}.

\subsection{Case Study}
\begin{figure}[!t]
\setlength{\abovecaptionskip}{0cm}
\includegraphics[scale=0.42]{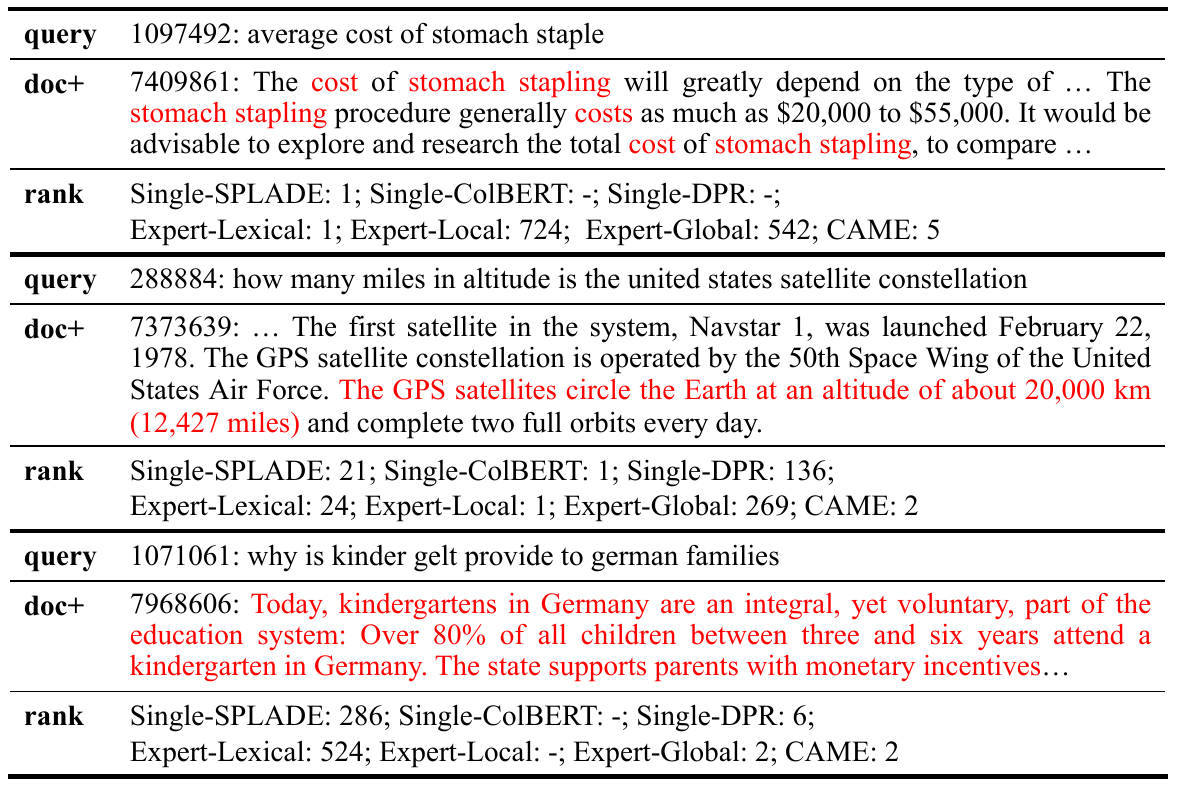}
\caption{Case study on MS MARCO. ``doc+'' denotes the relevant passage. ``Single-'' and ``Expert-'' denote a single-model retriever trained independently and the component expert in \modelname{} (SPLADE, ColBERT, and DPR are the counterparts of lexical, local, and global matching experts respectively). `-' in ranks means that it is beyond top 1000.}
\label{fig:case}
\end{figure}

In Figure~\ref{fig:case}, we show three representative cases in MS MARCO that have different relevance patterns to compare \modelname{}, its component experts, and their counterparts trained independently. 
We find that, for each case, only one or two models could rank the relevant passage at the top positions.
This supports our claim that different types of retrieval models are skilled at identifying different relevance patterns, and a single model could not handle all the cases.
By contrast, \modelname{} works well on all the three types of cases by leveraging the power of various types of retrieval experts.

We also compare the results of the component experts in \modelname{} (denoted as ``Expert-'') and their counterparts as single-model retrievers, and have some interesting observations:
(1) When there are clear disagreements between different types of retrievers, the retriever that ranks the ideal passage to top positions when trained individually performs similarly or even better when it is trained as a component in \modelname{}. The other single-model retrievers that are not good at the cases, by contrast, could have fluctuating performance when trained as experts in \modelname{}. This indicates that the joint training of the experts in \modelname{} with competitive learning mechanism can maintain or enhance their ability to identify their dedicated relevance patterns. 
(2) In the first two cases, although one component expert in \modelname{} ranks the relevant passage at top-1, the fusion results (i.e., \modelname{}) are not optimal. We impute it to the imperfect result fusion method. In the future, we need to explore better methods to fuse the retrieval results.

\section{Conclusion  and Future Work}
To fully exploit the capabilities of different types of retrieval models for effective retrieval, we propose a retrieval model based on the MoE framework and a novel competitive learning mechanism for the experts training.
Specifically, the MoE retriever consists of three representative matching experts, and the competitive learning mechanism develops and enhances their expertise on dedicated relevance patterns sufficiently with standardized and specialized learning. 
Empirical results on three retrieval benchmarks show that \modelname{} can achieve significant gains in retrieval effectiveness against the  baselines. 
In the future, we plan to further refine \modelname{} in terms of both effectiveness and efficiency:
(1) Using more advanced pre-trained models~\cite{zhou2022master,liu2022retromae} or exploring other result fusion methods to improve the performance;
(2) Investigating methods to improve the retrieval efficiency of \modelname{} without compromising its performance.

\bibliographystyle{ACM-Reference-Format}
\bibliography{paper}

\end{document}